\documentclass[english,12pt]{article}
\usepackage{array}
\usepackage{graphicx}
\usepackage{amssymb}
\usepackage{amsmath}
\usepackage{multirow}
\usepackage{prettyref}
\usepackage{babel}
\usepackage{units}
\usepackage[latin1]{inputenc}
\usepackage{amsfonts}
\usepackage{amssymb}
\usepackage{babel}
\usepackage{subcaption}
\usepackage[dvipsnames,usenames]{color}
\usepackage{cite}
\def\@fmsl@sh#1#2#3{\m@th\ooalign{$\hfil#1\mkern#2/\hfil$\crcr$#1#3$}}
 \def\eq#1\en{\begin{equation}#1\end{equation}}
\def\s[#1,#2]{[#1\stackrel{\star}{,}#2]}
\def\sx[#1,#2]{[#1\stackrel{\star_{x}}{,}#2]}

\newcommand{\D}{\nabla}

\def\gsim{\mathrel{\rlap{\lower4pt\hbox{\hskip1pt$\sim$}}

    \raise1pt\hbox{$>$}}}       %greater than or approx. symbol

\def\gsim{\mathrel{\rlap{\lower4pt\hbox{\hskip1pt$\sim$}}
    \raise1pt\hbox{$>$}}}       %greater than or approx. symbol

\begin{document}
\makeatletter
\def\fmslash{\@ifnextchar[{\fmsl@sh}{\fmsl@sh[0mu]}}
\def\fmsl@sh[#1]#2{%
  \mathchoice
    {\@fmsl@sh\displaystyle{#1}{#2}}%
    {\@fmsl@sh\textstyle{#1}{#2}}%
    {\@fmsl@sh\scriptstyle{#1}{#2}}%
    {\@fmsl@sh\scriptscriptstyle{#1}{#2}}}
\def\@fmsl@sh#1#2#3{\m@th\ooalign{$\hfil#1\mkern#2/\hfil$\crcr$#1#3$}}
\makeatother
%\baselineskip 24pt

%%%%%%%%%%%%%%%%%%%%%%%%%%%%%%%%%%%%%%%%%%%%%%%%%%%%%%%%%%%%%%%%%
%%%
%%%                      TITLE PAGE
%%%
%%%%%%%%%%%%%%%%%%%%%%%%%%%%%%%%%%%%%%%%%%%%%%%%%%%%%%%%%%%%%%%%%
\thispagestyle{empty}
\begin{titlepage}
\boldmath
\begin{center}
  \Large {\bf Singularity avoidance in quantum gravity}
    \end{center}
\unboldmath
\vspace{0.2cm}
\begin{center}
{{\large Iber\^e Kuntz$^{ab}$}\footnote{kuntz@bo.infn.it}, {\large Roberto Casadio$^{ab}$}\footnote{casadio@bo.infn.it}}
 \end{center}
\begin{center}
{\sl
$^a$
Dipartimento di Fisica e Astronomia, Universit\`a di Bologna,
\\
via Irnerio~46, I-40126 Bologna, Italy}
\\
$ $
\\
{\sl 
$^b$
I.N.F.N., Sezione di Bologna, IS - FLAG
\\
via B.~Pichat~6/2, I-40127 Bologna, Italy
}
\end{center}

\vspace{5cm}
\begin{abstract}
The purpose of this work is to investigate the consequences of quantum gravity for the singularity problem.
We study the higher-derivative terms that invariably appear in any quantum field theoretical model of gravity,
handling them both non-perturbatively and perturbatively.
In the former case, by computing the contributions of the additional degrees of freedom to a congruence of
geodesics, we show that the appearance of singularities is no longer a necessity.
In the latter, which corresponds to treating the quantised general relativity as an effective field theory,
we generalise the Hawking-Penrose theorems to include one-loop corrections of both massless matter and
graviton fluctuations.
\end{abstract}  
\end{titlepage}
%
%\pacs{}
%
%
%%%%%%%%%%%%%%%%%%%%%%%%%%%%%%%%%%%%%%%%%%%%%%%%%%%%%%%%%%%%%%%%
%%%
%%%                     INTRODUCTION
%%%
%%%%%%%%%%%%%%%%%%%%%%%%%%%%%%%%%%%%%%%%%%%%%%%%%%%%%%%%%%%%%%%%
%
\newpage
\section{Introduction}
\label{sec:intro}
Infinities in physics usually signal the existence of unknown phenomena taking place at new scales,
at which point one should replace our current theories (and possibly formalisms) for a more fundamental
description of Nature.
In general relativity, these infinities come in the form of singularities~\cite{HE} and they point to the breakdown
of our current understanding of gravity.
The standard view in the community is that quantum gravity should be able to resolve this issue by
smoothing out singularities.
Nonetheless, despite the many existing approaches to quantum gravity, there is no consensus
about what it is and how one should construct a quantum theory of spacetime, thus a proof of principle
for the singularity avoidance is yet to be found.
\par
Partial success has been obtained in the linear approximation, where one expands the metric around
some background vacuum solution and quantises only the vacuum fluctuation.
While this approach is background-dependent and not really a quantum theory of spacetime,
it allows for the quantisation of the gravitational field by employing standard methods of quantum
field theory.
Regardless of the gravitational bare action one starts with, one then invariably needs to deal with
higher-derivative terms to be able to handle ultraviolet (UV) divergences.
Whether these can be absorbed into a finite set of free parameters will however largely depend
on the choice of the bare action.
\par
In this work, we look at the singularity problem using two different models, both within the limits of
quantum field theory, and which differ from each other depending on whether we treat
higher-derivative terms perturbatively or non-perturbatively.
We first consider quadratic gravity, a higher derivative theory containing terms up to quadratic 
curvatures in the action and which has proven to be renormalisable~\cite{Stelle:1976gc} and
asymptotically free~\cite{Avramidi:1985ki}.
Quadratic gravity could in principle be seen as a fundamental theory of gravity if it was not for
the ghost present in its spectrum, i.e.~a particle with the wrong sign in front of its kinetic term,
which could lead to instabilities in the theory~\cite{Stelle:1976gc,Stelle:1977ry}
(see also Ref.~\cite{Salles:2018ccb} for a review).
But it can nonetheless be considered as a first approximation of a more fundamental theory
of quantum gravity.
\par
We will also explore quantum general relativity, treated as an effective field theory.
In this scenario, the theory is non-renormalisable~\cite{tHooft:1974toh} and
ghost-free~\cite{Simon:1990jn,Donoghue:1994dn,Burgess:2003jk}, the only degree of freedom
being the usual graviton.
The idea of effective field theories is to use the known degrees of freedom in the infrared (IR),
namely the graviton in our case, and compute radiative corrections to the interactions perturbatively
with respect to inverse powers of the Planck mass $M_p$.
It is very important to stress that this effective field theory does not constitute yet another approach
to quantum gravity.
It rather consists of a systematical study of quantum gravity in the IR,
regardless of what happens in the UV regime.
Any respectful UV completion of gravity, including quadratic gravity described in the last paragraph,
should produce the same results of the effective theory at low energies.
\par
To understand whether the Hawking-Penrose singularity theorems~\cite{HE} can be evaded in quantum gravity,
we make use of the celebrated Raychaudhuri equation~\cite{wald}~\footnote{As usual, we discard the contribution
due to the twist $\omega_{\mu\nu}$.}
\begin{equation}
\dot\theta = -\sigma_{\mu\nu}\sigma^{\mu\nu} -\frac13 \theta^2 - \Delta
\ ,
\label{eq:ray}
\end{equation}
which describes the divergence $\theta$ of a family of geodesics whose tangent vectors are $k^\mu$,
where $\sigma_{\mu\nu}$ is the shear tensor and
\begin{equation}
\Delta=R_{\mu\nu}k^\mu k^\nu
\label{eq:delta}
\end{equation}
is the discriminant (also known as Raychaudhuri scalar).
Since the first two terms in Eq. \eqref{eq:ray} are non-positive, the analysis of the convergence of
geodesics ultimately reduces to the study of the sign of $\Delta$.
Positive contributions to $\Delta$ (or negative contributions to $\dot\theta$) leads to focusing geodesics,
which reflects the attractive character of classical gravity.
Negative contributions to $\Delta$, on the other hand, would be interpreted as repulsive interactions,
which would be able to defocus geodesics, possibly preventing the formation of singularities.
\par
This paper is organized as follows. In Sec.~\ref{sec:higher}, we investigate higher-derivative gravity
by making the scalar and the ghost fields explicit in the action and studying their contribution to the
formation of singularities.
In Sec.~\ref{sec:qg}, we use the quantum action of general relativity to explore generalisations
of the Hawking-Penrose theorem at one-loop order.
We then discuss our findings in Sec.~\ref{sec:conc}.
\section{Quadratic gravity}
\label{sec:higher}
The divergence structure of general relativity at one-loop reveals the appearance of quadratic curvature
terms in the action.
The fact that the original Einstein-Hilbert action does not contain these terms indicates that general relativity
is non-renormalisable.
This motivated the inclusion of squared curvature terms in the bare action, leading to a modified theory of gravity.
This theory turns out to be renormalisable to all loop orders, but a ghost is present in the spectrum.
Several ideas have been proposed to get rid of this
ghost~\cite{Lee:1969fy,Cutkosky:1969fq,Tomboulis:1977jk,Tomboulis:1983sw,Kuntz:2019qcf}.
Here we stick to the position that the ghost is actually fictitious as it only appears as a byproduct of the
truncation of an action containing infinitely many terms~\cite{Kuntz:2019qcf}.
One can therefore project the ghost out by a suitable choice of the boundary conditions.
\par
The action of quadratic gravity reads
\begin{align}
	\label{eq:localaction}
        S= \frac{M_p^2}{2}\int d^4x \sqrt{-\tilde g} \left(\tilde R + \alpha \tilde R^2 
        + \beta \tilde R_{\mu\nu} \tilde R^{\mu\nu} + \gamma 
        \tilde R_{\mu\nu\rho\sigma}\tilde R^{\mu\nu\rho\sigma}\right)
        \ ,
\end{align}
where $\tilde R$, $\tilde R_{\mu\nu}$ and $\tilde R_{\mu\nu\rho\sigma}$ are the Ricci scalar, Ricci tensor and
Riemann tensor of the metric $\tilde g_{\mu\nu}$, respectively~\footnote{Note that the square of the Riemann
tensor is usually eliminated in favour of the other two curvature invariants by invoking Gauss-Bonnet theorem.
Here we choose to leave it explicit in the action just to follow the same notations commonly used in the literature.}.
The above action contains massive spin-0 and spin-2 fields in addition to the usual graviton (massless spin-2).
They can be made explicit via successive field redefinitions of the metric~\cite{Hindawi:1995an},
\begin{align}
\bar g_{\mu\nu} &= e^\chi\, \tilde g_{\mu\nu}
\\
g^{\mu\nu} &=\left[\det A(\phi_{\sigma\tau})\right]^{-1/2}\,\bar g^{\mu\alpha}\,A_{\alpha}^{\ \nu}
\ ,
\end{align}
such that the action~\eqref{eq:localaction} in the $g_{\mu\nu}$ frame reads
\begin{align}
        S
        =
        \frac{M_p^2}{2}\int \sqrt{-g}
        &
        \left\{
        R -\frac{3}{2} \left[A^{-1}(\phi_{\sigma\tau})\right]_\mu^{\ \nu}\,\nabla^\mu \chi\, \nabla_\nu \chi
        -\frac{3}{2}\, m_0^2\left[\det A(\phi_{\sigma\tau})\right]^{-1/2}\left(1-e^{-\chi}\right)^2
        \right.
        \nonumber
        \\
        &
        -g^{\mu\nu} \left[C^\lambda{}_{\mu\rho} 
        (\phi_{\sigma\tau}) C^\rho{}_{\nu\lambda} (\phi_{\sigma\tau}) - 
        C^\lambda{}_{\mu\nu} (\phi_{\sigma\tau}) C^\rho_{\rho\lambda}(\phi_{\sigma\tau})\right]
        \nonumber
        \\
        &
        \left.
        +\frac{1}{4}m_2^2 \left[\det A(\phi_{\sigma\tau})\right]^{-1/2}
        \left(\phi_{\mu\nu}\, \phi^{\mu\nu}-\phi^2\right)
        \right\}
        \ ,
\label{eq:newframeaction}
\end{align}
where $\phi\equiv g^{\mu\nu}\,\phi_{\mu\nu}$, 
$m_0 = (6\,\alpha + 2\,\beta + 2\,\gamma)^{-1/2}$ is the mass of the scalar field $\chi$ and
$m_2=(-\beta-4\,\gamma)^{-1/2}$ is the mass of the massive spin-2 particle $\phi_{\mu\nu}$.
The tensors $A_{\lambda}^{\ \nu}$ and $C^\lambda{}_{\mu\nu}$ are defined by
\begin{equation}
A_{\lambda}^{\ \nu}
=
\left(1+\frac{1}{2}\,\phi\right)\delta_{\lambda}^{\ \nu}
-\phi_{\lambda}^{\ \nu}\ .
\end{equation}
and
\begin{equation}
        C^\lambda{}_{\mu\nu}
        =
        \frac{1}{2} \left(\tilde g^{-1}\right)^{\lambda\rho}
        \left({\D}_\mu \tilde g_{\nu\rho} + {\D}_\nu \tilde g_{\mu\rho} - {\D}_\rho \tilde g_{\mu\nu}\right)
        \ ,
\end{equation}
where the connection $\D$ is compatible with the new metric $g_{\mu\nu}$ and $\tilde g_{\mu\nu}$
is seen as a function of $g_{\mu\nu}$.
The field $\phi_{\mu\nu}$ must satisfy the spin-2 consistent conditions
\begin{equation}
\label{eq:cond1}
\D^\mu\left(\phi_{\mu\nu} - g_{\mu\nu}\,\phi\right)
-
g^{\lambda\mu}\,C^\rho_{\ \lambda\mu}\left(\phi_{\rho\nu} - g_{\rho\nu}\,\phi\right)
-
g^{\lambda\mu}\,C^\rho_{\ \lambda\nu}\left(\phi_{\rho\mu} - g_{\rho\mu}\,\phi\right)
=
0
\end{equation}
and
\begin{equation}
\label{eq:cond2}
\phi
-m_2^{-2}
\left\{
\left[\det A(\phi_{\sigma\tau})\right]^{1/2}
\left[A^{-1}(\phi_{\sigma\tau})\right]_\mu^{\ \nu}\, \nabla^\mu \chi\, \nabla_\nu\chi
+
2\, m_0^2 \left(1-e^{-\chi}\right)^2
\right\}
=
0
\ .
\end{equation}
\par
The equations of motion are quite involved, thus we consider two different approximations.
First, let us assume that the massive spin-2 field $\phi_{\mu\nu}$ is isotropic, that is
\begin{align}
\phi_{\mu\nu} = \tfrac14\, \phi\, g_{\mu\nu}
\ ,
\label{eq:iso}
\end{align}
and the equations of motion for the metric reads
\begin{align}
R_{\mu\nu} - \frac12\, g_{\mu\nu}\,R
=
\
&
\frac32 \left(1+\frac{\phi}{4}\right)^{-1}
\left(\partial_\mu\chi\,\partial_\nu\chi
-
\frac12\, g_{\mu\nu}\,\partial_\rho\chi\,\partial^\rho\chi\right)
\nonumber
\\
&
+
\frac{3}{32}\left(1+\frac{\phi}{4}\right)^{-2}
\left(\partial_\mu\phi\,\partial_\nu\phi
-
\frac12\, g_{\mu\nu}\,\partial_\rho\phi\,\partial^\rho\phi
\right)
\nonumber
\\
&
-\frac32\, g_{\mu\nu}\, V(\chi,\phi)
\ ,
\label{eq:eomiso}
\end{align}
where
\begin{equation}
V(\chi,\phi)
=
\frac{m_0^2\left(1-e^{-\chi}\right)^2}
{2\left(1+\frac{\phi}{4}\right)}
+
\frac{m_2^2\, \phi^2}{16\left(1+\frac{\phi}{4}\right)^2}
\ .
\end{equation}
By trace-reversing Eq.~\eqref{eq:eomiso}, we obtain
\begin{align}
R_{\mu\nu}
=
\frac32 \left(1+\frac{\phi}{4}\right)^{-1}
\partial_\mu\chi\,\partial_\nu\chi
+
\frac{3}{32}\left(1+\frac{\phi}{4}\right)^{-2}
\partial_\mu\phi\,\partial_\nu\phi
+
\frac32\,g_{\mu\nu}\,V(\chi,\phi)
\ .
\end{align}
Contracting the above with a time-like vector $u^\mu$ gives the discriminant
defined in Eq.~\eqref{eq:delta} with $k^\mu=u^\mu$,
\begin{align}
\Delta
=
\left[
\frac32 \left(1+\frac{\phi}{4}\right)^{-1}
\partial_\mu\chi\,\partial_\nu\chi
+
\frac{3}{32}\left(1+\frac{\phi}{4}\right)^{-2}
\partial_\mu\phi\,\partial_\nu\phi
+
\frac32\,g_{\mu\nu}\,V(\chi,\phi)
\right]
u^\mu\, u^\nu
\ .
\end{align}
From the Penrose-Hawking theorems, singularities are unavoidable when $\Delta \geq 0$.
By defining the trajectory's proper time $u^\mu=d x^\mu/d\tau$,
we have $u^\mu\,\partial_\mu\chi=d\chi/d\tau\equiv \dot\chi$, and noting that on time-like
physical trajectories $g_{\mu\nu}\,u^\mu\,u^\nu= -1$, leads to
\begin{align}
\Delta
=
\frac32 \left(1+\frac{\phi}{4}\right)^{-1}\dot\chi^2
+
\frac{3}{32}\left(1+\frac{\phi}{4}\right)^{-2}\dot\phi^2
-
\frac32\,V(\chi,\phi)
\ .
\end{align}
Thus we conclude that singularities are no longer a clear necessity as we now have
negative contributions to $\Delta$.
\par
For example, singularities can be avoided in situations where the potential $V$ dominates
over the kinetic terms~\footnote{We are assuming $m_i^2>0$ ($i=0,2$) to avoid tachyonic instabilities.},
such as in the early universe.
In the vicinity of the Big Bang epoch, the potential is expected to dominate
so that inflation can take place, thus leading to the possible avoidance of the initial singularity.
Note also that singularities can be avoided even when the kinetic terms dominate due to the coupling
between the spin-0 and spin-2 fields if $\phi<-4$.
This should be confronted with Eq.~\eqref{eq:cond2}.
Upon using~\eqref{eq:iso}, we obtain
\begin{equation}
A_{\lambda}^{\ \nu}
=
\left(1+\frac{\phi}{4}\right)\delta_{\lambda}^{\ \nu}
\ ,
\end{equation}
and~\eqref{eq:cond2} will turn into a condition for $\chi$,
\begin{align}
m_2^2\,\phi
&
=
\left[\det A(\phi_{\sigma\tau})\right]^{1/2}
\left[A^{-1}(\phi_{\sigma\tau})\right]_\mu^{\ \nu}\, \nabla^\mu \chi\, \nabla_\nu\chi
+
2\, m_0^2 \left(1-e^{-\chi}\right)^2
\nonumber
\\
&
=
\left(1+\frac{\phi}{4}\right)
\partial_\mu\chi\,\partial^\mu\chi
+
2\, m_0^2 \left(1-e^{-\chi}\right)^2
\label{eq:inst}
\end{align}
or
\begin{align}
\phi
=
\frac{\partial^\mu\chi\partial_\mu\chi + 2m_0^2(1-e^{-\chi})^2}{m_2^2 - \frac14 \partial^\mu\chi\partial_\mu\chi}.
\end{align}
The condition $\phi<-4$ then translates into
\begin{align}
2\,m_0^2\,(1-e^{-\chi})^2 + 4\,m_2^2 > 0,
\end{align}
for $m_2^2<\tfrac14 \partial^\mu\chi\partial_\mu\chi$
(note that $\phi<-4$ has no solution for $m_2^2>\tfrac14 \partial^\mu\chi\partial_\mu\chi$).
In particular, the condition $\phi<-4$ is satisfied whenever the kinetic energy of $\chi$ dominates.
Let us see a concrete example of this. Consider an explicit isotropic metric of the form
\begin{align}
ds^2
=
-A\,dt^2+B\,dr^2+r^2\,d\Omega^2
\ ,
\end{align}
where $A$ and $B$ only depend on $r$ and $\chi$ is static. From Eq. \eqref{eq:inst}, we obtain
\begin{align}
m_2^2\,\phi
&
=
-\left(1+\frac{\phi}{4}\right)
\frac{(\chi')^2}{B}
+
2\, m_0^2 \left(1-e^{-\chi}\right)^2
\ .
\end{align}
Therefore, if $m_i^2>0$ ($i=0,2$), $\phi<-4$ requires that $B<0$.
Note that $B$ becomes negative, for example, inside the Schwarzschild radius.
This means that the defocusing of time-like geodesics is switched on as soon as they cross
the horizon, leading to the possible avoidance of the singularity at $r=0$ of a Schwarzschild black hole.
\par
Now let us investigate the scenario where $\phi_{\mu\nu}$ is not isotropic, but both massive fields are
seen as small perturbations around an arbitrary spacetime.
In this case, the action reads
\begin{align}
S = \int\mathrm{d}^4x\sqrt{-g}\left[\frac{M_p^2}{2}R + \mathcal{L}_\chi + \mathcal{L}_\phi \right],
\label{eq:actionpert}
\end{align}
where the spin-0 sector is given by
\begin{align}
        \mathcal{L}_\chi = -\frac{3}{2} \partial^\mu \chi \partial_\mu \chi 
        -\frac{3}{2} m_0^2(1-e^{-\chi})^2
\label{eq:s}
\end{align}
and the spin-2 sector reads
\begin{align}
\mathcal{L}_\phi
=
\ &
-\frac{1}{4} \left(\nabla_\mu \phi \nabla^\mu \phi 
-\nabla_\mu \phi_{\nu\rho} \nabla^\mu \phi^{\nu\rho}
+2 \nabla^\mu \phi_{\mu\nu} \D^\nu \phi - 2 \D_\mu \phi_{\nu\rho}\D^\rho \phi^{\mu\nu}\right) 
\nonumber
\\
&
+ \frac{1}{4} m^2_2 \left(\phi_{\mu\nu} \phi^{\mu\nu} -\phi^2\right)
\ .
\label{eq:pf}
\end{align}
The spin-2 conditions~\eqref{eq:cond1} and \eqref{eq:cond2} become
\begin{align}
\label{eq:condpert1}
\nabla^\mu\phi_{\mu\nu} = \phi=0
\ .
\end{align}
Varying Eq.~\eqref{eq:actionpert} with respect to $g^{\mu\nu}$ and imposing the conditions~\eqref{eq:condpert1},
leads to
\begin{align}
R_{\mu\nu} = 8\pi G\left(T_{\mu\nu}-\frac12 g_{\mu\nu}T\right), 
\end{align}
where $T_{\mu\nu} = T_{\mu\nu}^\chi + T_{\mu\nu}^\phi$ and
\begin{align}
8\pi G T_{\mu\nu}^\chi
&
=
g_{\mu\nu}\left[-\frac32 \partial_\rho\chi\partial^\rho\chi - \frac32 m_0^2 (1-e^{-\chi})^2\right] + 3\partial_\mu\chi\partial_\nu\chi,
\label{Tchi}
\\
8\pi G T_{\mu\nu}^\phi
&
=
\frac14 g_{\mu\nu}\left[\D_\rho\phi_{\alpha\beta}\D^\rho\phi^{\alpha\beta} + m_2^2 \phi_{\alpha\beta}\phi^{\alpha\beta}\right]
 2\D_\mu\phi_{\alpha\beta}\D_\nu\phi^{\alpha\beta}
\label{Tphi}
\end{align}
are the energy-momentum tensors of each field.
The discriminant is then given by
\begin{align}
\Delta
=
3\dot\chi^2 - \frac32 m_0^2 (1-e^{-\chi})^2 - \frac34 \D_\gamma\phi_{\alpha\beta}\D^\gamma\phi^{\alpha\beta}
+ \frac14 m_2^2 \phi_{\alpha\beta}\phi^{\alpha\beta} - 2(\D_0\phi^\alpha_{\ \beta})^2
\ .
\label{eq:deltapert}
\end{align}
Note that, while nothing can be said about the sign of the third term, we see that the first and fourth terms
are positive while the second and fifth terms are negative.
Therefore, as before, singularities can be avoided in situations where the negative terms are dominant.
We should stress that the negative contribution to $\Delta$ due to the last term in Eq.~\eqref{eq:deltapert}
corresponds to the fact that $\phi_{\mu\nu}$ is a ghost and thus mediates a repulsive interaction.
We can therefore ameliorate the singularity problem at the price of having ghost instabilities in the theory.
However, as it was shown in Refs.~\cite{Kuntz:2019qcf,Barnaby:2007ve,Kuntz:2019gup}, this ghost can
be projected out of asymptotic states by means of a contour in the Fourier space, which is justified as the
Ostrogradskian ghost is absent in a complete theory containing infinitely many diffeomorphism
invariants~\cite{Kuntz:2019qcf,Barnaby:2007ve}.
It should be noted that projecting out the ghost particle from the theory only prevents it from appearing
in external lines of Feynman diagrams, but it can still contribute to internal lines, which has precisely
the desired effect of mediating the repulsive interaction that leads to the last term in Eq.~\eqref{eq:deltapert}.
We should also mention that the energy conditions~\cite{HE,wald} appearing among the hypotheses of the singularity
theorems correspond to properties one expects for classical matter fields.
Hence, it is not unconceivable that they can be averted by the energy-momentum tensors~\eqref{Tchi} and~\eqref{Tphi}
given the purely quantum nature of the fields $\chi$ and $\phi_{\mu\nu}$. We note that similar conclusions to the above were recently reached in \cite{Kuipers:2019qby} using different methods.

Finally, we must confront our results with \cite{Li:2015bqa}, where it was proven that Ricci-flat spacetimes are exact solutions of general local and non-local higher derivative theories, indicating that singularities of Einstein's vacuum solutions should exist even beyond general relativity. In our approach, one must make a distinction between the Ricci tensor in the original frame $\tilde R_{\mu\nu}$ and the Ricci tensor in the transformed frame $R_{\mu\nu}$. Essentially, when one splits up the original metric $\tilde g_{\mu\nu}$ into the new background metric $g_{\mu\nu}$ and the additional massive fields $\{\chi,\phi_{\mu\nu}\}$, the curvature is also split into different pieces. One of these pieces is identified as the Ricci tensor of the new metric, while the others are the strengths corresponding to the massive fields $\chi$ and $\phi_{\mu\nu}$. For example, considering only the spin-2 sector, the relation between the Ricci tensor in different frames reads
\begin{equation}
\tilde R_{\mu\nu} = R_{\mu\nu} - \nabla_\mu C^\lambda_{\ \lambda\nu} + \nabla_\lambda C^\lambda_{\ \mu\nu} - C^\lambda_{\ \mu\rho}C^\rho_{\ \nu\lambda} + C^\lambda_{\ \mu\nu}C^\rho_{\ \rho\lambda}. 
\end{equation}
For $\phi_{\mu\nu}=0$, both the strength $C^\lambda_{\ \mu\nu}$ and the new Ricci tensor $R_{\mu\nu}$ vanish upon using the equation of motion, which implies $\tilde R_{\mu\nu}=0$. However, $\tilde R_{\mu\nu}=0$ does not imply $\phi_{\mu\nu}=0$. More generally, although the vanishing of the massive fields implies the vanishing of the original Ricci tensor $\tilde R_{\mu\nu}$, the opposite is not true. Therefore, Ricci flatness is not a preserved property under field redefinitions. If one then chooses to interpret the theory in the original frame, then the singularities of Einstein's vacuum solutions remain present as pointed out in \cite{Li:2015bqa}. But they are just a reflection of a bad choice of field variables. In fact, if one interprets the theory in the new frame, where gravity (described by the new metric $g_{\mu\nu}$) is inherently accompanied by two other fields, then the singularities can be unraveled. This seems to indicate a profound link between singularities and the choice of field variables, whose details will be worked out in the future.
\section{Quantum general relativity}
\label{sec:qg}
General relativity is known to be non-renormalisable, which until a few decades ago had been considered
as a long-standing problem since quantum general relativity would require an infinite number of observations
to make the theory predictive, should one proceed naively by using the old-fashioned methods
of quantum field theory.
However, within the realm of effective field theories, the non-renormalisability of general relativity is actually
a positive result as it shows the reason why classical general relativity is so successful in describing
the Universe as we see it:
new physics comes suppressed by the extremely huge Planck scale $M_p\sim 10^{19}\,$GeV.
\par
The idea of effective field theories is to organise all possible terms in the action in powers of $E/M$,
where $E$ is the typical energy of the problem under consideration and $M$ is a cutoff.
The zeroth order term is the basic action, the one which defines the degrees of freedom and
classical interactions.
Higher powers of $E/M$ contributes only to the latter, thus leading to corrections to vertices in
the Feynman diagrams, but not to the propagators.
For general relativity, one has $M\sim M_p$ and the bare action reads
\begin{align}
S= \int d^4x \sqrt{-g} \left(\frac{M_p^2}{2}R + \tilde b_1 R^2 + \tilde b_2 R_{\mu\nu} R^{\mu\nu}
+ \mathcal{O}(\partial^6 g)\right)
\ ,
\label{eq:effac}
\end{align}
where $\tilde b_i$ are bare coefficients.
The basic action is the Einstein-Hilbert action, which has the graviton as the only degree of freedom.
One must not confuse the actions~\eqref{eq:localaction} and \eqref{eq:effac}.
Although they look the same, they are treated differently and have different features.
In~\eqref{eq:localaction}, all terms are treated on the same foot, leading to a renormalisable theory
that contains other degrees of freedom besides the graviton.
The action~\eqref{eq:effac}, on the other hand, is the result of a non-renormalisable theory
(i.e.~general relativity) and must be treated perturbatively to comply with the effective field theory
approach, which is predictive at energies below $M_p$ and contains the graviton as the only particle
in the spectrum~\cite{Simon:1990jn,Donoghue:1994dn,Burgess:2003jk}.
\par
The quantisation of~\eqref{eq:effac} can be performed in the background field formalism, in which 
the metric is split as $g_{\mu\nu}=\bar g_{\mu\nu}+h_{\mu\nu}$, where $\bar g_{\mu\nu}$ is the
classical background metric, and the perturbation $h_{\mu\nu}$ is quantised.
Barvinsky~\textit{et al} developed a very general formalism to obtain the effective action $\Gamma$
of a gauge theory (including gravity) as an expansion in
curvatures~\cite{Barvinsky:1985an,Barvinsky:1987uw,Barvinsky:1990up,Vilkovisky:1984st}.
If we restrict to the cases where the only non-vanishing vacuum expectation value is the metric
$g_{\mu\nu}$ and only letting massless particles run in the loops, we obtain
\begin{align}
\Gamma
=
\int\mathrm{d}^4x\sqrt{-g}\bigg[
&
\frac{M_p^2}{2}R + b_1 R^2 + b_2 R_{\mu\nu}R^{\mu\nu}
+ c_1 R\log\left(\frac{-\Box}{\mu^2}\right)R + c_2 R_{\mu\nu}\log\left(\frac{-\Box}{\mu^2}\right)R^{\mu\nu}
\nonumber
\\
&
+ c_3 R_{\mu\nu\rho\sigma}\log\left(\frac{-\Box}{\mu^2}\right)R^{\mu\nu\rho\sigma}\bigg]
\ .
\label{eq:qaction}
\end{align}
Note that the coefficients of the local operators acquire a running $b_i = b_i(\mu)$ after renormalisation.
The coefficients of the non-local operators, on the other hand, are fully specified and depend solely on
the spin of the fields that have been integrated out.
\par

The equation of motion for the perturbation $h_{\mu\nu}$ around the background $\bar g_{\mu\nu}$
can be obtained in perturbation theory by evaluating the equation of motion from \eqref{eq:qaction} order by order in $h_{\mu\nu}$:
\begin{align}
&G_{\mu\nu}[\bar g] = 0,\label{eq:pert1}\\
&\frac{\delta G_{\mu\nu}[\bar g]}{\delta g^{\rho\sigma}}h^{\rho\sigma}+\Delta G_{\mu\nu}[\bar g] = 0,\label{eq:pert2}
\end{align}
where $G_{\mu\nu}$ is the Einstein tensor and $\Delta G_{\mu\nu}$ denotes the quantum corrections to Einstein's equation coming from the higher derivatives in \eqref{eq:qaction}.
At zeroth order, we obtain Eq. \eqref{eq:pert1} which is simply Einstein's equation for the background $\bar g_{\mu\nu}$. The leading order Eq. \eqref{eq:pert2}, on the other hand, gives precisely the equation of motion for $h_{\mu\nu}$:
\begin{equation}
\Box\bar h_{\mu\nu}
=
-\bar R\left[b_1+(c_1-c_3)\log\left(\frac{-\Box}{\mu^2}\right)\right]\bar R_{\mu\nu}
+ \frac{1}{2} \bar g_{\mu\nu} \bar R_{\rho\sigma}\left[b_2+(c_2+4c_3)\log\left(\frac{-\Box}{\mu^2}\right)\right]
\bar R^{\rho\sigma}
\ ,
\label{eq:prop}
\end{equation}
where $\bar h_{\mu\nu} = h_{\mu\nu} -\frac12 \bar g_{\mu\nu} h$, the bar is used for background objects
and the order in $h$ will be denoted by a superscript bracketed number.
\par
Let us start by looking at time-like congruences.
First, we expand the discriminant up to linear order,
\begin{equation}
\Delta = \bar R_{\mu\nu}k^\mu k^\nu + R^{(1)}_{\mu\nu}k^\mu k^\nu
\ ,
\label{eq:lineardisc}
\end{equation}
where $R^{(1)}_{\mu\nu} = -\frac12 \Box\bar h_{\mu\nu} + \frac14 \bar g_{\mu\nu}\Box \bar h$.
From~\eqref{eq:prop} and~\eqref{eq:lineardisc}, we obtain
\begin{align}
\Delta 
=
\
&
\bar R_{\mu\nu}k^\mu k^\nu
+ \frac12\bar R\left[b_1 + (c_1-c_3)\log\left(\frac{-\Box}{\mu^2}\right)\right]\bar R_{\mu\nu}k^\mu k^\nu
\nonumber
\\
&
+ \frac14\bar R\left[b_1+(c_1-c_3)\log\left(\frac{-\Box}{\mu^2}\right)\right]\bar R
- \frac14 \bar R_{\rho\sigma}\left[b_2+(c_2+4c_3)\log\left(\frac{-\Box}{\mu^2}\right)\right]\bar R^{\rho\sigma}
\ .
\label{eq:tldisc}
\end{align}
Given that the background metric satisfies Einstein's field equations in the absence of a cosmological constant,
i.e.~$\bar R_{\mu\nu} = 0$, all terms in Eq.~\eqref{eq:tldisc} vanish and we obtain a generalisation
of the Hawking-Penrose theorem for singularities along time-like geodesics at one-loop order.
This is a direct consequence of the fact that the right hand side of Eq.~\eqref{eq:prop} vanishes
for a Ricci flat spacetime and it obviously extends to null-like trajectories as we will confirm below.
Our finding is in agreement with the results of Refs.~\cite{Calmet:2017qqa,Calmet:2019eof}
for the Schwarzschild black hole.
\par
Note, however, that the vanishing of the discriminant $\Delta$ up to one-loop order only indicates
that nothing can be said regarding the formation of singularities at this order.
Since we are working in perturbation theory, $\Delta=0$ represents a marginal result and it is therefore
inconclusive.
One must go beyond one-loop corrections to study the sign of $\Delta$.
Likewise, within perturbation theory, $\Delta$ is clearly dominated by the classical contribution
$\bar R_{\mu\nu}k^\mu k^\nu$, thus the sign of $\Delta$ is not changed by the loop contributions
unless the tree level result is marginal, i.e.~$\bar R_{\mu\nu}k^\mu k^\nu=0$.
\par
We should also stress two important points regarding this generalisation.
First, the background can be completely arbitrary and need not be described by an Einstein manifold. In this case, one would obtain an additional contribution on the right-hand side of Eq. \eqref{eq:prop} given by $-G_{\mu\nu}[\bar g]$. We chose an Einstein background to simplify our analysis.
Secondly, even if it is described by Einstein's equations (as we assumed above), the background is Ricci flat only in the absence
of a cosmological constant and for vanishing matter vacuum expectation values, in which case
the macroscopic energy-momentum tensor is zero.
We have indeed assumed that all background matter fields are zero in order to obtain the
quantum action~\eqref{eq:qaction}.
In the presence of a non-zero cosmological constant $\Lambda$, for example, we can have both
positive and negative contributions according to the sign of the coupling constants $c_i$,
which ultimately depend on the spin of the integrated particles, and $b_i(\mu)$,
whose sign is dictated by their renormalisation group.
\par
With the above points in mind, we can now state the one-loop generalisation of the Hawking-Penrose
theorem for time-like congruences.
Let us suppose that $\bar R_{\mu\nu}k^\mu k^\nu=0$, otherwise the standard Hawking-Penrose result
holds.
Then, a necessary (but not sufficient) condition for the avoidance of time-like singularities is that
\begin{equation}
\bar R\left[b_1 + (c_1-c_3)\log\left(\frac{-\Box}{\mu^2}\right)\right]\bar R
- \bar R_{\rho\sigma}\left[b_2+(c_2+4c_3)\log\left(\frac{-\Box}{\mu^2}\right)\right]\bar R^{\rho\sigma}
> 0
\ .
\end{equation}
\par
Observe that the quadratic theory studied in Sec.~\ref{sec:higher}, when treated perturbatively,
should agree with the results of quantum general relativity at low energies.
In fact, treating higher-order curvature terms as perturbations corresponds to handling the fields
$\chi$ and $\phi_{\mu\nu}$ perturbatively as well.
One can then compare both approaches order by order.
For example, the scalar field in Eq.~\eqref{eq:newframeaction} is defined by
\begin{equation}
\chi = \log(1+3\,m^2 R) = 3\,m^2 R + \mathcal O(R^2)
\ ,
\end{equation}
thus the mass term $\chi^2\sim R^2$ reproduces the square of the Ricci scalar in Eq.~\eqref{eq:tldisc}.
This in fact corresponds to putting the field $\chi$ on shell (at tree level), which makes physical sense
because the masses of $\chi$ and $\phi_{\mu\nu}$ are supposedly of the order of the Planck mass,
thus both fields decouple from the theory at energies below $M_p$.
\par
Let us now look at null-like congruences.
From $\bar g_{\mu\nu}k^\mu k^\nu = 0$, we find
\begin{equation}
\Delta
=
\bar R_{\mu\nu}k^\mu k^\nu 
+ \frac12 \bar R\left[b_1 + (c_1 - c_3)\log\left(\frac{-\Box}{\mu^2}\right)\right]\bar R_{\mu\nu}k^\mu k^\nu
\ ,
\label{eq:nldisc}
\end{equation}
and we again conclude that the Hawking-Penrose theorem is fulfilled for a Ricci flat spacetime in the case
of null-like singularities as it should be expected from Eq.~\eqref{eq:prop}.
Nonetheless, for null-like vectors $k^\mu$, the discriminant $\Delta$ in Eq.~\eqref{eq:nldisc} vanishes even
in the presence of a cosmological constant because the Ricci tensor $\bar R_{\mu\nu}$ is proportional
to the metric $\bar g_{\mu\nu}$.
Non-trivial contributions to $\Delta$ are only possible when either classical matter is present or on
non-instanton backgrounds.
Unfortunately, as discussed earlier, one-loop corrections are only sizeable when $\bar R_{\mu\nu}k^\mu k^\nu=0$,
making all terms in Eq.~\eqref{eq:nldisc} vanish and leading to $\Delta=0$.
As we explained before, this means that the study of formation of singularities along null-like geodesics is inconclusive
at one-loop order and one must go beyond this approximation in order to be able to determine the sign of $\Delta$.
\section{Conclusions}
\label{sec:conc}
In this paper, we have shown how quantum corrections to gravity can make the singularity problem less
severe by giving positive contributions to $\dot\theta$ via the Raychaudhuri equation.
\par
We first considered the fourth-derivative extension of general relativity, which is renormalisable but contains
a ghost field in the Lagrangian, i.e.~a field whose kinetic term has a negative norm, and which can be interpreted
as a repulsive force.
It is precisely this repulsive feature of the ghost, together with the potential terms, that could be able to prevent
the formation of singularities in the spacetime.
On the other hand, the very same feature is also responsible for vacuum instabilities in the theory should it be
present in asymptotic states. 
We argued that one can take advantage of the repulsive character of the ghost without facing instability issues
by projecting the ghost particle out of the asymptotic spectrum. We also found out theoretical evidence suggesting the existence of an interesting link between singularities and the choice of field variables. The details of this new finding will be investigated in a future project.
\par
In the second part of the paper, we looked at the problem from an effective field theory perspective,
thus treating the action perturbatively as an expansion in inverse powers of the Planck mass.
We showed that the Hawking-Penrose theorems can be generalised to include one-loop corrections
when the background is Ricci flat, but the conditions for the formation of singularities are modified otherwise.
Within perturbation theory, the tree level contribution naturally dominates over the loop corrections,
thus the one-loop correction will dictate the fate of the discriminant only in the marginal case in which the classical
contribution vanishes.
\par
In the present work we have tried to reach general conclusions by analysing the actions~\eqref{eq:localaction}
and \eqref{eq:qaction} without considering specific physical systems.
It will be interesting to further study this issue by employing explicit cosmological models or descriptions of the 
gravitational collapse of compact objects.
\section*{Acknowledgments}
I.K.~and R.C.~are partially supported by the INFN grant FLAG.
The work of R.C.~has also been carried out in the framework of activities of the National Group of Mathematical Physics
(GNFM, INdAM) and COST action Cantata.
%
% 
%%%%%%%%%%%%%%%%%%%%%%%%%%%%%%%%%%%%%%%%%%%%%%%%%%%%%%%%%%%%%%%%%
%%%
%%%                     BIBLIOGRAPHY
%%%
%%%%%%%%%%%%%%%%%%%%%%%%%%%%%%%%%%%%%%%%%%%%%%%%%%%%%%%%%%%%%%%%%
%
%

\end{document}